\documentclass[a4paper,fleqn,usenatbib,useAMS]{mnras}
\usepackage{xcolor}
\usepackage{graphicx}
\usepackage{pdflscape}	
\usepackage{amsmath}	
\usepackage{amssymb}	
\usepackage{multicol}        
\usepackage{bm}		
\usepackage{ascii}
\usepackage{float} 
\usepackage{caption} 
\usepackage{ulem} 

\usepackage{ulem}

\DeclareGraphicsExtensions{.png}

\definecolor{grassgreen}{RGB}{0, 186, 0}

\definecolor{orange}{RGB}{255, 140, 0}

\newcommand {\be} {\begin{equation}}
\newcommand {\ee} {\end{equation}}

\newcommand{\msun}{~M$_{\odot}$}

\newcommand{\hii}{\ifmmode \text{H}\,\textsc{ii} \else H~{\scshape ii}\fi\xspace}
\newcommand{\Ha}{\ifmmode \text{H}\alpha \else H$\alpha$\fi\xspace}
\newcommand{\Hb}{\ifmmode \text{H}\beta \else H$\beta$\fi\xspace}
\newcommand{\oi}{\ifmmode [\text{O}\,\textsc{i}] \else [O~{\scshape i}]\fi\xspace}
\newcommand{\Oi}{\ifmmode [\text{O}\,\textsc{i}]\lambda 6300 \else [O~{\scshape i}]$\lambda 6300$\fi\xspace}
\newcommand{\oii}{\ifmmode [\text{O}\,\textsc{ii}] \else [O~{\scshape ii}]\fi\xspace}
\newcommand{\Oii}{\ifmmode [\text{O}\,\textsc{ii}]\lambda 3726 + \lambda 3729 \else [O~{\scshape ii}]$\lambda 3726 + \lambda 3729$\fi\xspace}
\newcommand{\nii}{\ifmmode [\text{N}\,\textsc{ii}] \else [N~{\scshape ii}]\fi\xspace}
\newcommand{\Nii}{\ifmmode [\text{N}\,\textsc{ii}]\lambda 6584 \else [N~{\scshape ii}]$\lambda 6584$\fi\xspace}
\newcommand{\oiii}{\ifmmode [\text{O}\,\textsc{iii}] \else [O~{\scshape iii}]\fi\xspace}
\newcommand{\Oiii}{\ifmmode [\text{O}\,\textsc{iii}]\lambda 5007 \else [O~{\scshape iii}]$\lambda 5007$\fi\xspace}
\newcommand{\sii}{\ifmmode [\text{S}\,\textsc{ii}] \else [S~{\scshape ii}]\fi\xspace}
\newcommand{\Sii}{\ifmmode [\text{S}\,\textsc{ii}]\lambda 6716 + \lambda 6731 \else [S~{\scshape ii}]$\lambda 6716 + \lambda 6731$\fi\xspace}
\newcommand{\wha}{\ifmmode W_{\mathrm{H}\alpha} \else $W_{\mathrm{H\alpha}}$\fi\xspace}
\newcommand{\ww}{\ifmmode W2 - W3 \else $W2 - W3$\fi\xspace}
\newcommand{\nuvr}{\ifmmode NUV - r \else $NUV - r$\fi\xspace}

\newcommand{\starlight}{\textsc{starlight}\xspace}

\defcitealias{CidFernandes2010}{CF10}
\defcitealias{CidFernandes2011}{CF11}
\defcitealias{Kewley2001}{K01}
\defcitealias{Stasinska2006}{S06}

\usepackage[T1]{fontenc}
\usepackage{ae,aecompl}

\usepackage[utf8x]{inputenc}

\title[Why do many early-type galaxies lack emission lines?]{Why do many early-type galaxies lack emission lines? I. Fossil clues}

\author[Herpich et al.]
{ F.\ Herpich$^{1,\,2}$\thanks{E-mail: herpich@astro.ufsc.br}, 
  G.\ Stasi\'nska$^{3}$,
  A.\ Mateus$^{1}$, 
  N. Vale Asari$^{1,4,5}$,
  R. Cid Fernandes$^{1}$
 \\
  $^{1}$Departamento de F\'{\i}sica--CFM, Universidade Federal de Santa Catarina, C.P.\ 476, 88040-900, Florian\'opolis, SC, Brazil \\
  $^{2}$Instituto de Astronomia, Geof\'{\i}sica e Ci\^{e}ncias Atmosf\'{e}ricas, Universidade de S\~{a}o Paulo, R. do Mat\~{a}o 1226, 05508-090 S\~{a}o Paulo, Brazil \\
  $^{3}$LUTH, Observatoire de Paris, PSL, CNRS, UMPC, Univ Paris Diderot, 5 place Jules Janssen, 92195 Meudon, France\\
  $^{4}$School of Physics and Astronomy, University of St Andrews, North Haugh, St Andrews KY16 9SS, UK\\
  $^{5}$Royal Society--Newton Advanced Fellowship\\
}

\date{Accepted 2018 August 30. Received 2018 August 30; in original form 2018 April 18}

\pubyear{2018}

\begin{document}

\label{firstpage}
\pagerange{\pageref{firstpage}--\pageref{lastpage}}
\maketitle

\begin{abstract}
Early-type retired galaxies (RGs, i.e. galaxies which no longer form stars) can be divided into two classes: those with no emission lines, here dubbed lineless RGs, and those with emission lines, dubbed liny RGs. Both types of galaxies contain hot low-mass evolved stars (HOLMES) which emit ionizing photons. The difference must thus lie in the presence or absence of a reservoir of ionizable gas. From a volume-limited sample of 38\,038 elliptical galaxies, we explore differences in physical properties between liny and lineless using data from the SDSS, WISE and GALEX catalogues. To avoid biases in the comparison, we pair-match liny and lineless in stellar-mass, redshift and half-light Petrosian radius. We detect marginal differences in their optical stellar ages and NUV luminosities, indicating that liny RGs have an excess of intermediate-age (0.1--5 Gyr) stellar populations. Liny RGs show higher dust attenuation and $W3$ luminosities than their lineless counterparts. We also find that the amount of warm gas needed to explain the observed \Ha luminosity in liny RGs is $10^5$--$10^8$\msun, and that their \nii/\oii emission-line ratios are typical of those of the most massive star-forming galaxies. Taken together, these results rules out the following sources for the warm gas in liny RGs: mass-loss from intermediate-mass stars, mergers with metal-poor galaxies and intergalactic streams. They imply instead an inflow of enriched gas previously expelled from the galaxy or a merger with a metal-rich galaxy. The ionization source and the origin of the gas producing the emission lines are thus disconnected.
\end{abstract}


\section{Introduction}\label{sec:intro}

Until recently the text-book vision of early-type galaxies (ETGs) was that they are composed mostly of old stars and almost devoid of cold gas and dust. However, already five decades ago it was realised that a non negligible fraction of them contained some gas, either neutral \citep[e.g.][]{Balkowski1972AA21303B,Gallagher1975ApJ2027G} or ionized \citep{Mayall1958IAUS523M,Osterbrock1960ApJ132325O}. It was also pointed out \citep[e.g.][]{Tinsley1972ApJ178.319T} that continuous mass-loss from old stars could lead to the formation of younger generations of stars in early-type galaxies. The far ultraviolet excess observed in the central region of M31 \citep{Code1969PASP81.848C}, first tentatively attributed to hot, highly evolved stars \citep{Hills1971AA12.1H}, was later suggested to arise from  hot young main-sequence stars \citep{1971AA15.403T}. Various ionization sources for the emission-line gas observed in elliptical galaxies were considered, such as decay of turbulence or ultraviolet radiation from hot stars \citep{Osterbrock1960ApJ132325O}.

In 1980, \citet{Heckman1980AA87.152H} performed a systematic spectroscopic survey of the nuclei of nearby galaxies, and found that, while the nuclei of spiral galaxies often had spectra similar to \hii regions, those of elliptical galaxies looked like the scaled-down version of the spectra of active galactic nuclei (AGN), with low luminosity and low excitation (but different from that of \hii regions). He called those objects LINERS, for `low ionization nuclear emission-line regions'. A wealth of hypotheses were proposed to explain the spectra of LINERs, such as nuclear activity \citep[e.g.][]{Ho1996ApJ.462.183H,Barth2001ApJ.546.205B,Sabra2003ApJ.584.164S,Filippenko2003ASPC.290.369F,Chiaberge2005ApJ.625.716C,Ricci2015MNRAS.451.3728R}, shocks \citep{Filippenko1984ApJ.285.458F,Ho1993ApJ.417.63H}, young, massive stars \citep[e.g.][]{Filippenko1992ApJ.397L.79F,Shields1992ApJ.399L.27S,Ho1993ApSS.205.19H,Ho1996ApJ.462.183H,Colina1997ASPC.113.477C,Alonso-Herrero2000ApJ.530.688A,Eracleous2002ApJ.565.108E,Rampazzo2005AA.433.497R}, or evolved low-mass stars \citep[e.g.][]{Stasinska2008,Annibali2010AA.519A.40A,Capetti2011AA.529A.126C,Yan2012,Yan2013IAUS.295.328Y,Singh2014IAUS.304.280S,Hsieh2017ApJ.851L.24H}. There is still no consensus. The LINER problem has become one of the most discussed topics concerning galaxies with emission lines. In the 1990s, much progress had been achieved in the theoretical evolutionary synthesis of stellar populations, and some works described  the spectral evolution of coeval stellar populations until very late stages where low mass stars reached the white dwarf stage \citep{Bruzual1983ApJ273105B,Bruzual1993ApJ.405.538B,Fioc1997ASSL210.257F}. \citet{Binette1994} then showed that the observed emission-line spectra in elliptical galaxies (for which data were still very scarce at that epoch) could well be explained by the ionization of hot, low-mass evolved stars (dubbed HOLMES\footnote{We prefer the denomination of HOLMES to that of `post-AGB stars' since in the stellar evolution community the latter term has for the last thirty years been used exclusively to designate the late stage of stellar evolution \textit{previous} to the planetary nebula phase \citep[see][]{vanWinckel2003ARAA41.391V}.} in later studies; e.g. \citealt{Flores-Fajardo2011MNRAS415.2182F} and \citealt{CidFernandes2011}, hereafter CF11).

In the 2000s, the Sloan Digital Sky Survey \citep[SDSS; ][]{York2000}, which obtained imaging photometry and optical spectra for nearly one million galaxies, revolutionised the field of galaxy studies. At the same time, techniques were developed to analyse the stellar populations of the galaxies and extract the pure emission-line spectra from the observed one which included information on both the stellar populations and the interstellar gas \citep{CidFernandes2005,Roig2015ApJ808.26R,Morelli2016MNRAS463.4396M}. This allowed for the first time a correct determination of the emission-line spectra of ETGs and a panoramic view of the galaxy spectra in emission-line diagnostic diagrams \citep{Kauffmann2003a}. 

In the \oiii/\Hb vs \nii/\Ha plot (the so-called BPT diagram after \citealt*{Baldwin1981}) the emission-line galaxies were cleanly separated between a `star-forming' (SF) wing and a wing containing hosts of active galactic nuclei (AGN). The lower part of the AGN wing was inappropriately labelled LINER \citep{Kauffmann2003a,Kewley2006}, creating some confusion: in the SDSS, spectra were obtained through a 3 arcsec fibre, and for objects at redshifts larger than about 0.03, covered much more than the sole nucleus. Later, the denomination LIER \citep{Belfiore2016MNRAS461.3111B} was proposed to underline that the spectra did not necessarily concern galaxy nuclei. In 2008, \citet{Stasinska2008} showed that ionization by the HOLMES inferred from the stellar population studies of the observed SDSS galaxies could well explain the LINER-type emission-line ratios observed for ETGs. In 2011, \citetalias{CidFernandes2011} proposed a diagnostic diagram (\wha vs \nii/\Ha, dubbed the WHAN diagram) to distinguish true LINERs from galaxies ionized by their HOLMES (the so-called retired galaxies introduced by \citealp{Stasinska2008}, meaning `retired from their star-forming activity').

2D spectroscopy, now available thanks to integral field units such as CALIFA and MaNGA (\citealt{Sanchez2012AA538A.8S} and \citealt{Bundy2015ApJ798.7B}, respectively) then confirmed that, in many cases, the WHAN diagram correctly distinguished retired galaxies from galaxies hosting a LINER in their nucleus \citep{Sarzi2010,Singh2013,Gomes2015arXiv151100744G}.

Presently, the widely accepted view of ETGs is that they contain an interstellar medium containing dust \citep{Goudfrooij1995AA.298.784G}, neutral hydrogen with masses of $10^6$--$10^9$\msun\ \citep[e.g. ][]{Krumm1979ApJ.228.64K,Goudfrooij1994MNRAS.271.833G,Oosterloo2010MNRAS.409.500O,Serra2012MNRAS.422.1835S,Lagos2014MNRAS.443.1002L,Woods2014MNRAS439.2351W}, molecules \citep{Combes2007MNRAS.377.1795C,Kaviraj2012MNRAS.423.49K,Davis2015MNRAS.449.3503D}, and have haloes of hot gas \citep{Sarzi2013MNRAS.432.1845S}. They also contain emission-line zones extending out to several kiloparsecs \citep[e.g.][]{Sarzi2006MNRAS.366.1151S,Singh2013,Woods2014MNRAS439.2351W,Gomes2015arXiv151100744G,Belfiore2017MNRAS.466.2570B}. HOLMES are one of the preferred options for the ionization of this gas. 

This new and broadly consensual vision of the ETGs seems now to have occulted the old picture and perhaps diminished the interest in those ETGs that \textit{do not present emission lines}. Some notable exceptions are the work by \citet{Rudnick2017ApJ.850.181R} who studied a sample of ETGs with and without \oii emission at redshifts between 0.4 and 0.8, and that by \citet{Belfiore2017MNRAS.466.2570B}, who analysed a sample of ETGs with MaNGA, looking for the reason behind the presence of gas and dust in some of these galaxies. As shown by \citetalias{CidFernandes2011} and \citet{Stasinska2015MNRAS}, such objects constitute about half of the whole population of ETGs. They were dubbed \textit{passive galaxies} (PG) in \citetalias{CidFernandes2011}, and \textit{lineless retired} (LLR) in \citet{Herpich2016MNRAS4621826H}. Here we will call them `lineless' RGs, by opposition to retired galaxies with emission lines, which we will call `liny' RGs. 

As already shown by \citetalias{CidFernandes2011}, lineless and liny RGs have almost identical distribution of many physical parameters, such as (1) stellar mass; (2) optical colors; (3) stellar populations mean age and metallicity; etc. So how is it that such a large proportion of ETGs does not show emission lines at all? In this series of two papers we try to understand what makes a retired galaxy lineless or liny. More precisely, given that the stellar populations of both types of galaxies are very similar and have been shown to be able to produce a low-level ionization of their gaseous content, why is it that some RGs do not show any emission lines? In this paper, we look at the fossil properties of liny and lineless ETGs, meaning properties linked to their star formation history or to the chemical enrichment of their emission-line gas. In a companion paper (Mateus et al., in prep.), we will discuss  environmental clues.  We will attack our study by considering the information provided by three large surveys: SDSS in the optical, WISE in the mid-infrared and GALEX in the UV.

This paper is organised as follows: Section \ref{sec:datasamples} presents the data and sample selection; Section \ref{sec:diff} presents a differential analysis of the stellar properties of the two classes of RGs that are the subject of this work; Section \ref{sec:gas} discusses the origin of  the gas in liny RGs; Section \ref{sec:discussion} puts together our findings; and finally Section \ref{sec:conclusions} summarises our results.

Throughout this work  we adopt a flat cosmological model with $\Lambda$-$\mathrm{CDM}$ with $H_0 = 70\,\mathrm{km\,s^{-1}Mpc^{-1}},\ \Omega_M = 0.3$  and $\Omega_{\Lambda} = 0.7$.

\section{Data and samples}\label{sec:datasamples}

We use the publicly available data from three large surveys, covering a spectral range from the ultraviolet to the infrared (SDSS, WISE and GALEX; \citealt{York2000,Wright2010,Martin2005}). We also take advantage from two other projects: the spectral synthesis of stellar populations with the \starlight code \citep[]{CidFernandes2005} and the morphology analysis from the Galaxy Zoo \citep{Lintott2008MNRAS}. In the following, we describe the surveys and explain our sample selection.

\subsection{The data}\label{sec:surveys}

We build our sample with the observations of the Sloan Digital Sky Survey\footnote{\url{http://www.sdss.org/}}. This survey took photometric and spectroscopic measures of a fourth of the sky in the visible range.  The 7th Data Release \citep[DR7;][]{Abazajian2009} has photometric observations of about 350 million objects in five bands ($ugriz$), with effective wavelengths of 3551, 4686, 6165, 7481 and 8931\,\AA, respectively.  SDSS DR7 has spectra  of more than a million galaxies in the range from 3800 to 9200\,\AA, with mean spectral resolution of $\lambda/\Delta\lambda\sim1800$ and obtained through fibres of 3\arcsec\ in diameter.

The Wide-Field Infrared Survey Explorer (WISE\footnote{\url{http://www.nasa.gov/mission_pages/WISE/main/index.html}}), observed the entire sky in four bands of the mid-infrared, $W1$, $W2$, $W3$ and $W4$, with central wavelengths 3.4, 4.6, 12 and 22\,$\mu$m, respectively. The All-Sky Data Release has positional data for more than 500 million objects with signal-to-noise $SN > 5$ to at least one band. The photometric calibration of WISE was made using the Vega system, and the conversion factor from Vega to AB is 2.683, 3.319, 5.242 and 6.604 for $W1$, $W2$, $W3$ and $W4$, respectively. We use this calibration to calculate WISE luminosities. However, as we do not directly compare WISE to SDSS magnitudes, we report WISE magnitudes in the Vega system.

The Galaxy Evolution Explorer (GALEX) observed the whole sky in two ultraviolet photometric bands. Its 6th General Release (GR6) contains over 200 million photometric sources with measures in the far- and near-ultraviolet \citep[$FUV$ and $NUV$, respectively; ][]{Martin2005}. The $FUV$ band covers a spectral range from 1344 to 1786\,\AA, with a central wavelength of 1528\,\AA. The $NUV$ band covers from 1771 to 2831\,\AA, with a central wavelength of 2310\,\AA\ \citep{Bianchi2014AdSpR}. The GALEX mission performed a variety of surveys,  the main one being the All Sky Survey (AIS), which covers an area of 26\,000 square degrees with a magnitude limit of $m_\mathrm{AB} = 20.5$.

\subsection{The sample}\label{sec:sample}

We start from a sample of 926\,246 galaxies from SDSS-DR7 which have been analysed with the stellar population synthesis code \starlight\footnote{\url{http://starlight.ufsc.br}}. From those, we select only objects from SDSS Main Galaxy Sample, with half-light surface brightness $\mu_{50} \leqslant 24.5\,\mathrm{mag\ arcsec^{-2}}$, and absolute $r$ magnitude $M_r \leqslant -20.43$. This gives us 553\,316 objects.  We match this sample to the MPA-JHU value-added catalogue to use their `\textsc{spectofibre}' parameter (i.e. to correct the SDSS-DR7 spectral flux calibration for extended sources, see \url{http://www.mpa-garching.mpg.de/SDSS/DR7/raw_data.html}).  This sample is then matched to the Galaxy Zoo\footnote{\url{http://www.galaxyzoo.org} contains the morphological information for almost 900 thousand galaxies of the SDSS-DR7 \citep{Lintott2008MNRAS,Lintott2011MNRAS}.}  catalogue, from which we obtain 515\,320 galaxies.

In order to study retired ETGs removing red spirals or blue spirals with a `retired' bulge, we select only those objects in the Galaxy Zoo catalogue for which the probability of being elliptical is higher than 0.5, and keep only the RGs (i.e. objects having $W_{\Ha} < 3$\,\AA\ which is the criterion defined by \citetalias{CidFernandes2011}), obtaining 168\,786 objects.

We then remove all the objects for which the SDSS spectrum is not `clean' in the region of the \Ha line (e.g. containing bad pixels or sky lines) as described in \citet{Herpich2016MNRAS4621826H} to guarantee that our lineless RG sample does not contain galaxies that might have unnoticed emission lines. We impose a signal-to-noise cut of $SN_{\Ha} > 3$ to liny RGs to select only those with reliable line measurements\footnote{This means that 12\,277 galaxies cannot be stated to be either a liny or a lineless RG.}. We thus obtain 59\,662 liny RGs ($0.5 \leqslant W_{\Ha} < 3$\,\AA) and 96\,844 lineless RGs ($W_{\Ha} < 0.5$\,\AA).

We then pair match the two RG samples in mass ($M_\star$), redshift ($z$) and half-light radius at $r$ band ($R_{50}$), finding which liny best matches a given lineless RG. After excluding a few outliers (as detailed in Appendix \ref{app:pairing}), we find 96\,836 liny--lineless pairs. The mean absolute difference between pairs is $|\Delta \log M_\star| = 0.007$\,dex, $|\Delta z| = 0.001$ and $|\Delta R_{50}| = 0.06$\,kpc, with standard deviations of $0.018$\,dex, $0.003$ and $0.19$\,kpc, respectively.

Finally, we select a volume limited sample in the redshift range $0.04 < z < 0.095$, which contains 24\,575 liny--lineless pairs, composed of 38\,038 retired galaxies (i.e.\ a liny RG can be a match to more than one lineless RG). This is our pair-matched optical Volume Limited Sample (VLS).  We then define two other sub-samples: (1) the infrared sample using WISE data ($SN_{W3} > 3$; WVLS), with 6944 pairs; and (2) the near-ultraviolet sample using the $NUV$ band from GALEX ($SN_{NUV} > 3$; GVLS), with 8121 pairs. These three samples will be used throughout this work.

\subsection{Stellar population analysis}

The comparisons of liny and lineless RGs presented below are partly based on direct measurements and partly on properties derived from our \starlight\ spectral fits. The latter were carried out using a set of 150 simple stellar populations (SSPs) from \citet{Bruzual2003MNRAS.344.1000B}, spanning 6 metallicities from $Z = 0.005$ to $2.5 Z_\odot$, and 25 ages from 1 Myr to 18 Gyr. The models use ``Padova 1994" evolutionary tracks and a \citet{Chabrier2003PASP.115.763C} initial mass function. Extinction was modelled as due to a foreground screen following a Galactic reddening law with $R_V=3.1$ \citep{Cardelli1989ApJ.345.245C}. This same setup was used in, for instance, \citet{Mateus2006MNRAS.370.721M}, \citet{Mateus2007MNRAS.374.1457M}, \citet{ValeAsari2009MNRAS.396L.71V}, and \citetalias{CidFernandes2011}.

Of the properties derived from this analysis we will focus on the (fibre-corrected) stellar mass $M_\star$\footnote{$M_\star$ is the total stellar mass of the galaxy corrected for aperture effects by using the photometric $z$-band fiber-to-total flux ratio, as in  \citet{Kauffmann2003a}. We have verified that using the $r$-band instead leads to the same masses to within $\pm 0.05$ dex. This extrapolation assumes no significant mass-to-light variations from the inner 3 arcsec to the galaxy as a whole, a hypothesis which is not critical for ETGs as those in our sample (\citealt{Gonzalez-Delgado2015AA.581A.103G}; Garcia-Benito et al. 2018, submitted).}, the stellar extinction $A_V$, the luminosity weighted mean stellar age $\langle t_\star \rangle$\footnote{$\langle t_\star\rangle$ is actually defined as $ \equiv 10^{\langle\log t_\star\rangle}$, where $\langle\log t_\star\rangle$ is the mean log age weighted by the contributions of different populations to the flux at normalisation wavelength $\lambda_0 = 4020$ \AA\ (see equation 2 of \citealt{CidFernandes2005}).}, and the predicted rate of ionizing photons produced by populations older than $10^8$ yr, $Q_{\rm HOLMES}$, as computed from the $\lambda < 912$ \AA\ model spectra. 
 
Uncertainties in these properties range from those related to noise in the data and the inversion method itself to others related to the ingredients (evolutionary tracks and spectral libraries) used in the analysis. \citet{CidFernandes2014AA.561A.130C} examines these issues in detail. Though the setup is not identical to the one used in our fits, their overall conclusions stand valid.

A key aspect of our study is that our comparison of liny and lineless systems is differential. By focusing on the differences in the stellar population properties among pairs of galaxies in these two families, we largely mitigate the effects of potential biases and uncertainties in the analysis. It is also worth noting that experiments with more up-to-date SSP models confirm the results reported here (Werle 2018, in prep.).

Given our interest in the ionizing radiation field, it is relevant to point out that we deliberately ignore the predicted Lyman continuum from populations younger than $10^8$ yr. This is done for several reasons. First, the ionizing photons output from young stars using spectral synthesis in the optical range is quite uncertain. The number of O stars in the stellar libraries we use is of just a few for all spectral types and metallicities, while the number of ionizing photons from individual O stars varies by almost two orders of magnitudes \citep{Martins2005AA.436.1049M}, making any interpolation difficult. In addition, the signatures of O stars in the optical range appear only in the blue part of the spectrum and are not very strong \citep{Walborn1990PASP.102.379W}. Finally they mostly rely on  hydrogen  and helium lines. The spectral region of hydrogen lines is removed from the synthesis done by STARLIGHT since it is affected also by emission from the interstellar medium. Therefore, the ionizing budget of young stellar populations cannot be derived accurately (a typical error of a factor of 2--4 is likely). Second, an optically insignificant contribution from $t \la 10^7$ yr populations may well dominate the predicted $\lambda < 912$ \AA\ flux, rendering the resulting predictions highly uncertain. For our galaxies we typically find that young stars account up to just 1.5\% of the flux at 4020 \AA, but this irrelevant component is enough to make them $\sim 3$ times brighter than old stars at ionizing energies. Third, imperfect modeling of hot phases in the evolution of low mass stars (like the horizontal branch) as well as blue stragglers (and possibly other effects associated to binary stars, e.g., \citealt{Eldridge2008MNRAS.384.1109E}) may well lead spectral fitting codes like \starlight\ to use young populations to compensate for the blue emission missing from old ones. This is what \citet{Ocvirk2010ApJ.709.88O} referred to as the `fake young bursts' problem, where a genuinely old population is best fit by a mixture of an old and a young one due to these effects (see also \citealt{Koleva2008MNRAS.385.1998K}; \citealt{CidFernandes2010MNRAS.403.780C}).

\section{Differential analysis of the two retired groups}\label{sec:diff}

\begin{figure}
\centering
\includegraphics[width=.47\textwidth]{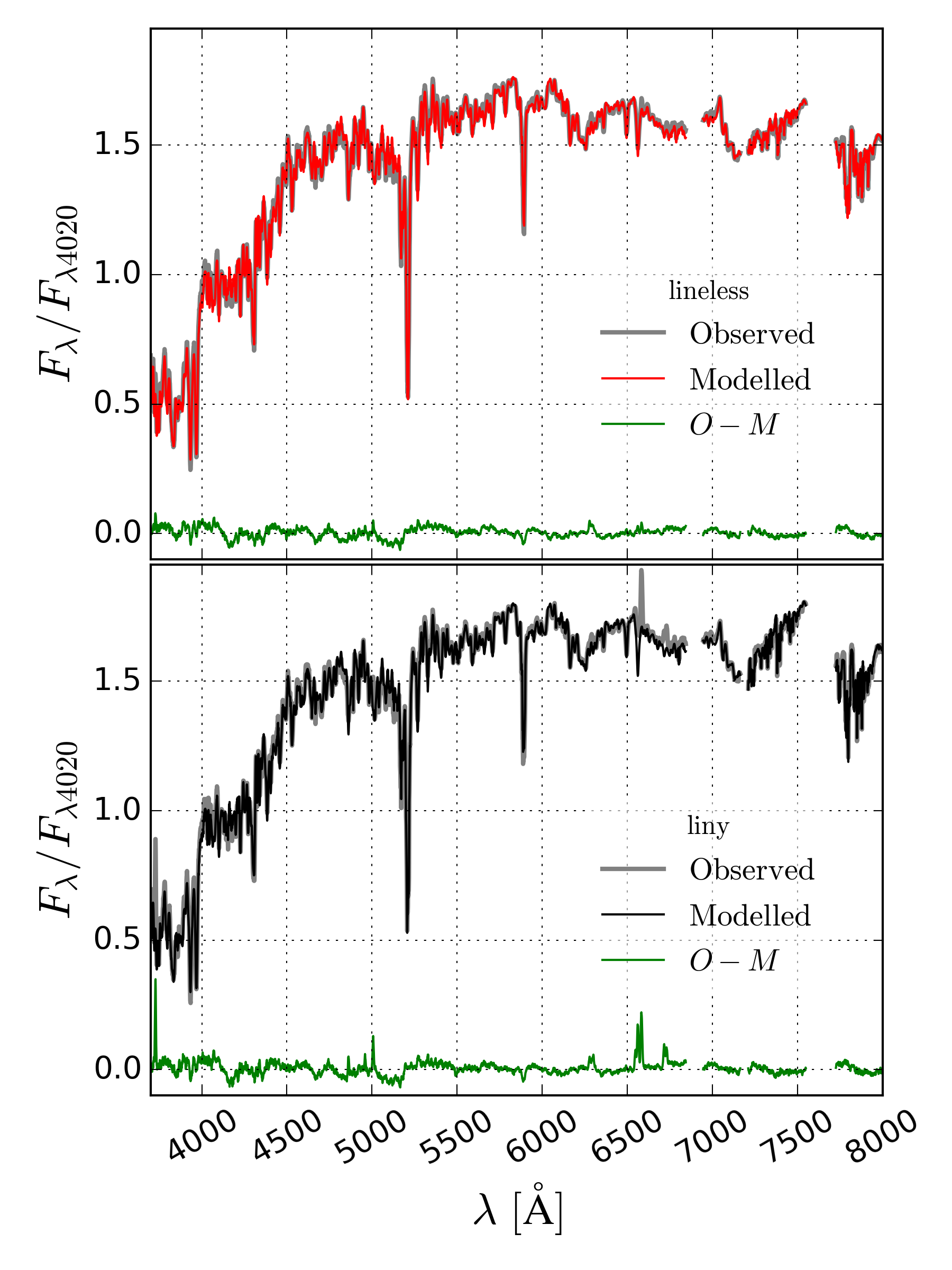}
\caption{Upper panel: Observed lineless stacked spectrum using 102 lineless spectra (gray), the stacked model from \starlight (red) and the residual spectrum observed minus modelled ($O - M$, green). Bottom panel: same as upper panel but for liny.}
\label{fig:synthesis}
\end{figure}

The two galaxy classes considered in this work are very similar, at least in the optical. And by construction our paired sample has galaxies which are even more similar. We show an example of the spectral features of liny and lineless RGs in Fig.~\ref{fig:synthesis}. Each panel shows the observed stacked spectrum of lineless and liny RGs, as well the fit from the spectral synthesis and the observed minus model residual spectrum. The stacking was made selecting galaxies in a bin near the mean stellar mass ($\log\ M_\star/M_\odot = 11.16 \pm 0.10$), redshift ($z = 0.0705 \pm 0.0025$) and Petrosian radius ($R_{50} = 5.2 \pm 1.0$ arcsec) of our paired sample. This selection leaves us with 102 lineless--liny pairs; the bin limits were deliberately chosen to select little no more than a hundred pairs. The stacked spectra change very little regardless of the limits chosen.

\subsection{Liny and lineless retired galaxies look the same\dots }\label{sec:equal}

\begin{figure}
\centering
\includegraphics[width=.47\textwidth]{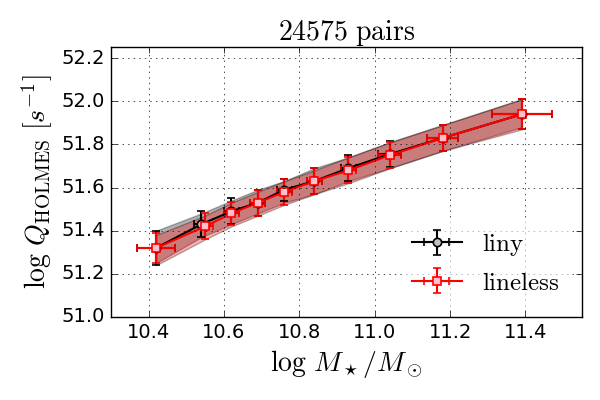}
\caption{$Q_\mathrm{HOLMES}$, the rate of ionizing photons produced by stellar populations older than $10^8$ yr versus the stellar mass of the galaxy. Liny RGs are in black and their matched lineless RGs in red. The markers are the median value for bins of mass containing approximately the same number of objects. The shadowed regions represent the quartiles of each distribution. The bars are the median absolute deviation to the median.}
\label{fig:qholmes}
\end{figure}

Fig.~\ref{fig:qholmes} compares $Q_{\rm HOLMES}$, the total rate of hydrogen ionizing photons produced by HOLMES, for liny and lineless RGs paired as explained in Section \ref{sec:sample}. The values of $Q_{\rm HOLMES}$ are obtained from the \starlight decomposition of the galaxy spectra, counting only the photons produced by stellar populations older than $10^8$ yr. In each category, the galaxies are grouped by stellar mass, $M_\star$, in bins of approximately the same number of objects. The curves are the median values of $Q_{\rm HOLMES}$ for each $M_\star$-bin as a function of $M_\star$. The shadowed zones represent the quartiles of each distribution and the bars are the values for the median absolute deviation. Quantities in black refer to liny RGs and those in red to lineless RGs. In this plot, the black and red features cannot be distinguished, indicating that RGs of same stellar mass, observed at the same redshift, and with the fibre covering the same amount of galaxy light, produce exactly the same number of ionizing photons per second due to HOLMES, whether they show emission lines or not. This is due to the fact that, for a canonical stellar initial mass function, the ionizing photon production rate of HOLMES does not depend much on the detailed star formation history for look-back times larger than $10^8$ yr \citepalias[see figure 2 of][]{CidFernandes2011}. Thus liny and lineless RGs are not different in terms of the ionizing power of their HOLMES.

\begin{figure}
\centering
\includegraphics[width=.47\textwidth]{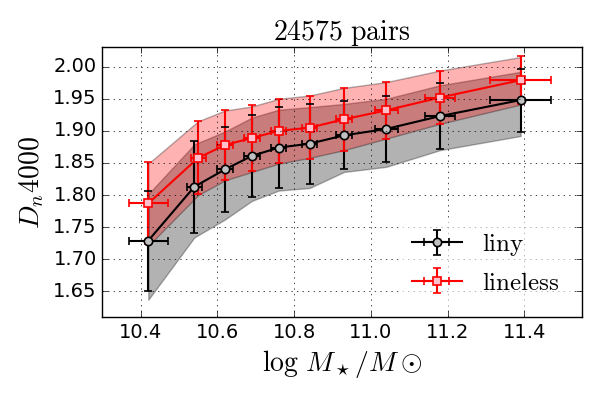}
\caption{
Same as Fig.~\ref{fig:qholmes} but for the $D_n4000 \mathrm{\AA}$ break.
}
\label{fig:d4000}
\end{figure}

\begin{figure}
\centering
\includegraphics[width=.47\textwidth]{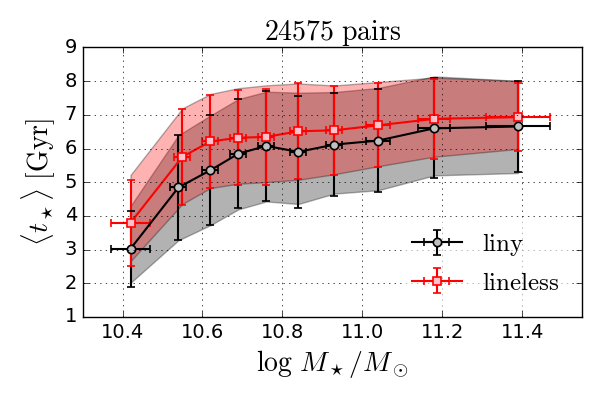}
\caption{
As Fig.~\ref{fig:qholmes} but for the mean stellar age weighted
by light.
}
\label{fig:tstar}
\end{figure}

It is interesting to compare other aspects of the star formation histories of these two classes of galaxies. One is provided by the $D_n4000$ index \citep{Balogh1999ApJ.527.54B}. This index is dependent on the mean age  of the stellar population and -- to a lesser extent -- on their metallicities \citep{Poggianti1997AA325.1025P}, with larger values occurring for galaxies with older stellar populations. Fig.~\ref{fig:d4000} shows that liny RGs tend to have slightly smaller values of $D_n4000$, at the verge of significance. Fig.~\ref{fig:tstar} compares the flux-weighted stellar age $\langle t_\star\rangle$ as obtained by \starlight between the two families. Here again liny RGs tend to show slightly smaller mean stellar ages. The effect is very small, but perceptible, contrary to the case of $Q_{\rm HOLMES}$, and occurs at all galaxy masses. These smaller ages cannot be due to ongoing star-formation, otherwise these galaxies would have a higher $W_{\Ha}$ and SF-like line ratios, contrary to what is observed. Instead, this result points to an excess of intermediate-age populations in liny with respect to lineless RGs.

An interesting feature  in Fig.~\ref{fig:d4000} and \ref{fig:tstar} is that the differences in the stellar population ages between liny and lineless RGs tend to decrease toward larger galaxy masses.  As we will see through this paper, this can be related with the limited amount of material that produces the emission lines. In such case, the effects produced either by the gas \textit{per se}, or by a younger stellar population, are masked by the rest of the galaxy and become easier to notice in less massive ETGs. In summary, liny and lineless RGs do not differ in their output of ionizing photons produced by HOLMES. They however seem to differ, on average, in their stellar populations of intermediate ages, but only slightly.

\begin{figure}
\centering
\includegraphics[width=.47\textwidth]{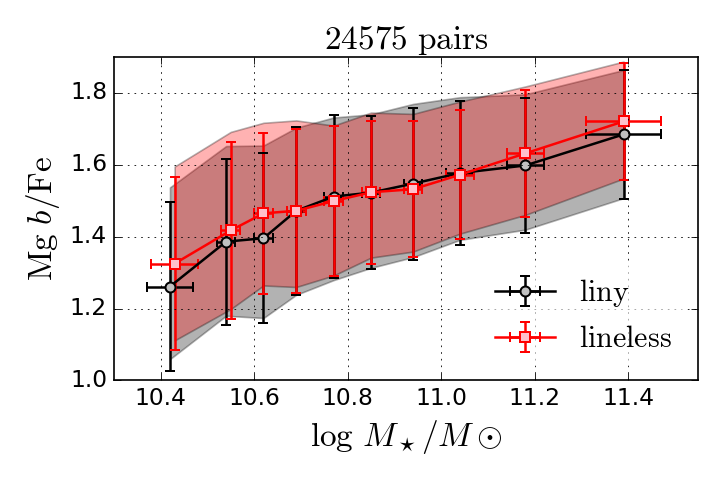}
\caption{
As Fig.~\ref{fig:qholmes} but for the $\alpha$/Fe index Mg$\,b$/Fe where Fe stands for the mean flux of Fe$\lambda\lambda5270,5335$.
}
\label{fig:licks}
\end{figure}

It is interesting to test whether these differences are somehow related to differences in $\alpha$/Fe ratios among the two classes. Our SSP base models do not account for this effect, but empirical indices like Mg$\,b$/Fe \citep{Thomas2003MNRAS.339.897T} can be used to tackle this issue. We have collected measurements of both Mg$\,b$ and Fe\footnote{Fe stands for the mean value of the flux for the $\lambda5270$ and $\lambda5335$ lines.} from the MPA/JHU catalog (\url{https://wwwmpa.mpa-garching.mpg.de/SDSS/DR7/SDSS\_indx.html}) and compared the alpha/Fe-sensitive Mg$\,b$/Fe ratio of liny and lineless. Fig.~\ref{fig:licks} shows the results and no difference is found.

\subsection{\dots but there are differences: dust}\label{sec:dust}

\begin{figure}
\centering
\includegraphics[width=.45\textwidth]{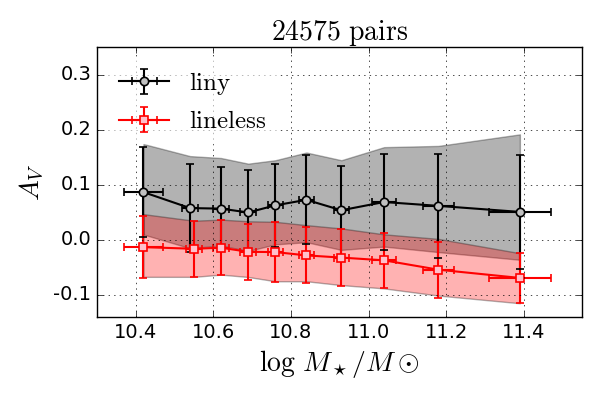}
\caption{
As Fig.~\ref{fig:qholmes} but for the attenuation obtained with the \starlight $A_V$.
}
\label{fig:AV}
\end{figure}

We now turn to indicators of the presence of dust. An obvious one is, of course, dust extinction. From \starlight, it is possible to estimate the extinction of stellar light by interstellar grains, assuming that dust is distributed in a slab between the stars and the observer. This is of course not the situation encountered in ETGs.  However, since the two families of galaxies considered are similar, it makes sense to investigate possible differences in $A_V$. Fig.~\ref{fig:AV} compares the values of $A_V$ between the two families of RGs as a function of $M_\star$. Lineless RGs have values of $A_V$ close to zero, while liny RGs have significantly higher values with a median value between 0.06 and 0.1. Negative values, encountered especially for lineless RGs, are obviously not physical. As discussed in \citet{CidFernandes2005}, our spectral fits allow for $A_V < 0$ to account for uncertainties in the flux calibration and Galactic extinction, as well as those associated with the modeling of dust attenuation itself\footnote{The code allows for $A_V > -1$, but values bellow $-0.2$ occur in less than 3\% of our sample.}. It is worth noting that negative values of the extinction are also obtained without \starlight, as in the study by \citet{Kauffmann2003a}. More importantly, imposing $A_V \ge 0$ does not erase the difference in $A_V$ between liny and lineless RGs, which is the key result here.
 
Using Occam's razor argument, we may expect the extinction law to be similar for both families of galaxies. In that case, Fig.~\ref{fig:AV} indicates a larger amount of dust in the line of sight for liny galaxies. Qualitatively, this is precisely what is expected since liny galaxies contain warm gas which is likely mixed with some dust. Note that the estimated $A_V$ does not necessarily allow a measure of the abundance of dust, since it was obtained under the hypothesis that the dust forms a screen between the stars and the observer.

\begin{figure}
\centering
\includegraphics[width=.45\textwidth]{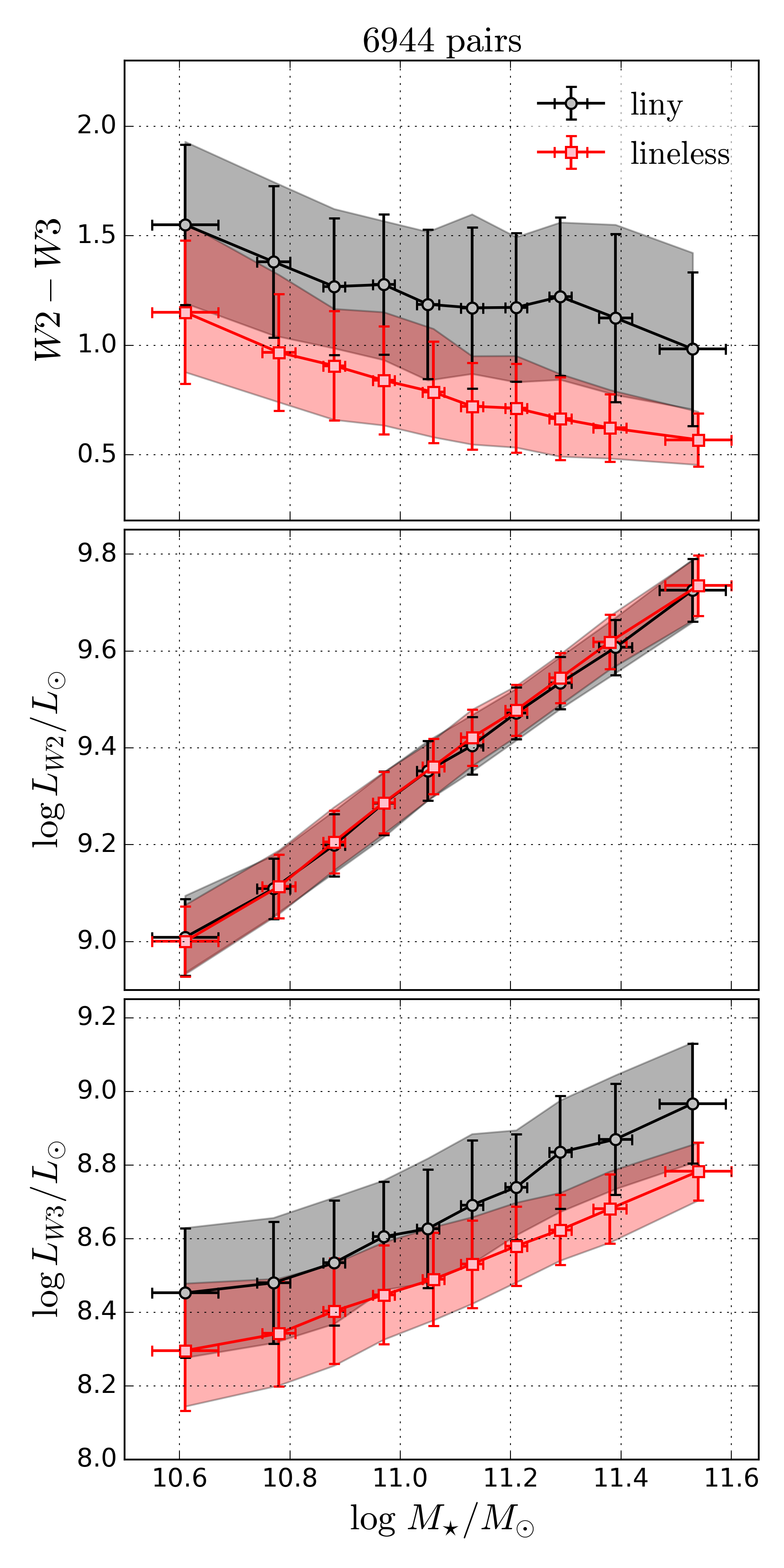}
\caption{
As Fig.~\ref{fig:qholmes} but comparing the WISE $W3 - W2$ colour (top panel); $W2$ and $W3$ luminosities (middle and bottom panels, respectively).
}
\label{fig:w2w3}
\end{figure}

Mid-infrared data, on the other hand, are able to directly detect the presence of hot dust. The $W3$ filter in WISE contains the Polycyclic Aromatic Hydrocarbon (PAH) 11.2\,$\mu$m and 12.7\,$\mu$m features, which are greatly enhanced in the presence of the hard UV field able to excite the PAH grains \citep{Draine2007}. Fig.~\ref{fig:w2w3} compares WISE data for the two RG families. The top panel shows the behaviour of the $W2 - W3$ colour. Liny RGs are clearly `redder' than lineless ones. As the two bottom panels show, this difference in color is due to an excess in the W3 band. Such an excess is generally interpreted as a sign of recent star formation, where the radiation field provided by the hot young stars  excites the surrounding PAHs \citep{Draine2007,daCunha2008,Vermeij2002,Peeters2004ApJ,Wu2005,Xu2015ApJ808159X}. But it could also be due to the presence of stars which have left the asymptotic giant branch and heat their expelled dust shells \citep{Justtanont1996,Cassara2013MNRAS4362824C,Villaume2015ApJ}.

\subsection{Other differences: UV }\label{sec:UV}

\begin{figure}
\centering
\includegraphics[width=.45\textwidth]{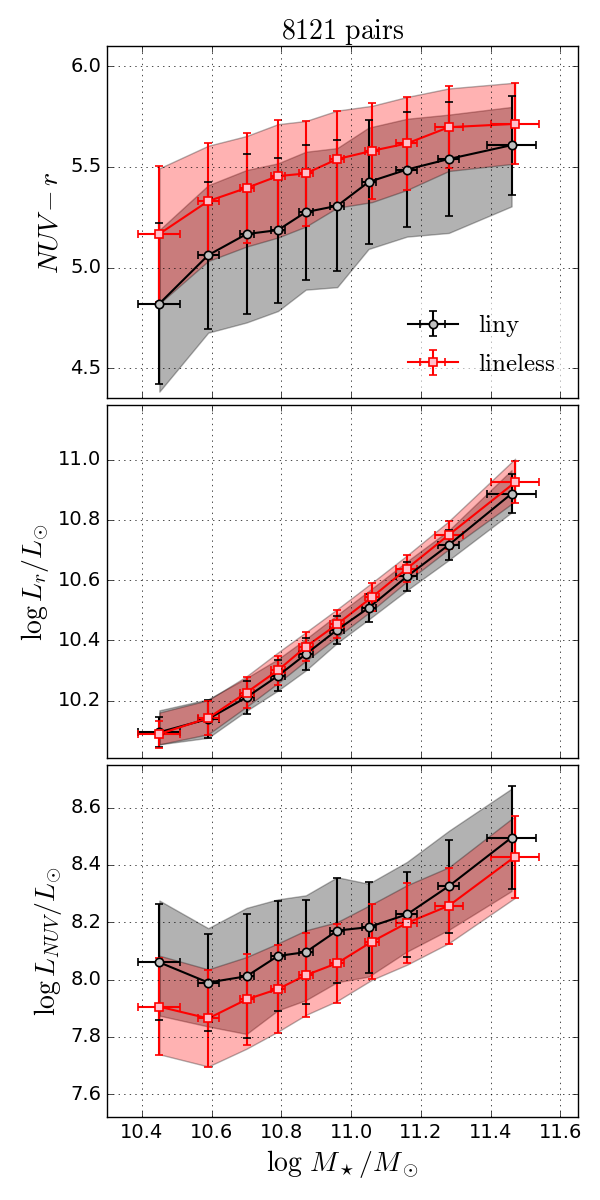}
\caption{
As Fig.~\ref{fig:qholmes} but for the $NUV - r$ colour (top panel); $L_r$ and the $L_{NUV}$ (middle and bottom panel, respectively).
}
\label{fig:UV}
\end{figure}

Ultraviolet data from GALEX can also give a clue on the nature of the stellar populations. The upper panel of Fig.~\ref{fig:UV} shows the behaviour of the $NUV-r$ colour. Liny RGs are clearly bluer than lineless ones and, as the bottom two panels show, this is entirely due to the luminosity in the NUV band. Because of their larger $A_V$'s (Fig.~\ref{fig:AV}), this stronger UV emission in linys becomes even more evident applying extinction corrections.

This indicates an excess population of intermediate age stars (0.1--5 Gyr) in liny RGs\footnote{An excess of HOLMES could also enhance the ultraviolet emission but from Fig.~\ref{fig:qholmes} the HOLMES populations are indistinguishable between liny and lineless RGs.} (see also \citealt{Kaviraj2007ApJS173619K,Kaviraj2011MNRAS411,Ko2013ApJ.767.90K,Young2014MNRAS.444.3408Y}). We note however that those stars are probably not the ones which ionize the emitting gas, as has sometimes been argued to reinforce the interpretation that a blue $NUV-r$ colour indicates ongoing star formation \citep{Yi2005ApJ619L.111Y}. Indeed, in all the liny RGs considered here, the ionizing radiation from the HOLMES is sufficient to explain the observed \Ha luminosities. The ultraviolet difference observed here therefore does not trace the source of ionization (HOLMES), but the radiation emitted by middle-age main-sequence stars.

\section{The nature of the emitting gas }\label{sec:gas}

The nature of the emitting gas in liny RGs is an important clue for what makes a retired galaxy liny or lineless.  At first sight, one could think that the gas naturally comes from the stellar winds of intermediate mass stars, not only from previous stages of the HOLMES which provide the present-day ionizing radiation, but also that of stars which have not yet reached the stage of HOLMES and those which were initially more massive than the progenitors of present-day HOLMES.  

The mass $M_{\Ha}$ of gas needed to produce the \Ha luminosity
in liny RGs can be obtained by combining the two equations: 
\begin{equation}\label{eq:lhalpha} 
L_{\Ha} = \int n(H^+)\, n_e\, \varepsilon_\mathrm{H\alpha}\, dV,
\end{equation}
where $L_{\Ha}$ is the \Ha luminosity,  $n(H^+)$ and $n_e$ are the H$^+$ and electron densities,  $\varepsilon_{\Ha}$ is the \Ha emissivity,
and
\begin{equation}\label{eq:mass}
M_{\Ha} = \int 1.4\, n(H^+)\, m_H\, dV,
\end{equation}
where $m_H$ is the  mass of an hydrogen atom. We obtain 
\begin{equation}\label{eq:masstohalpha}
\frac{M_{\Ha}}{L_{\Ha}} = \frac{1.4 m_H}{\langle n_e\rangle \varepsilon_{\Ha}}.
\end{equation}
where $\langle n_e\rangle$ is an averaged electron density, which is not well known. Using $\langle n_e\rangle = 0.1\, \mathrm{cm^{-3}}$, we find $M_{\Ha} \sim 10^8\,M_\odot$. If we use the value found by \citet{Johansson2016MNRAS461.4505J} from stacked RG galaxy spectra, $\langle n_e\rangle = 100\,\mathrm{cm^{-3}}$, this mass becomes $M_{\Ha} \sim 10^5\,M_\odot$. This value is similar to that quoted by \citet{Davis2016MNRAS.457.272D} for local ETGs and by \citet{Pandya2017ApJ.837.40P} for the most massive ETGs using the same $n_e \sim 100\,\mathrm{cm^{-3}}$ density. Clearly the gas supplied by mass-loss from evolved stars  (which is about the same as the present-day mass in stars) is more than enough to explain the observed \Ha luminosities. This fact was already noted by \citet{Belfiore2017MNRAS.466.2570B}. Other options can be external gas from a recent merger or cooling filaments from galaxy haloes \citep[e.g.][]{Lagos2014MNRAS.443.1002L,Lagos2015MNRAS.448.1271L,Davis2016MNRAS.457.272D}.

\begin{figure*}
\centering
\includegraphics[width=.9\textwidth]{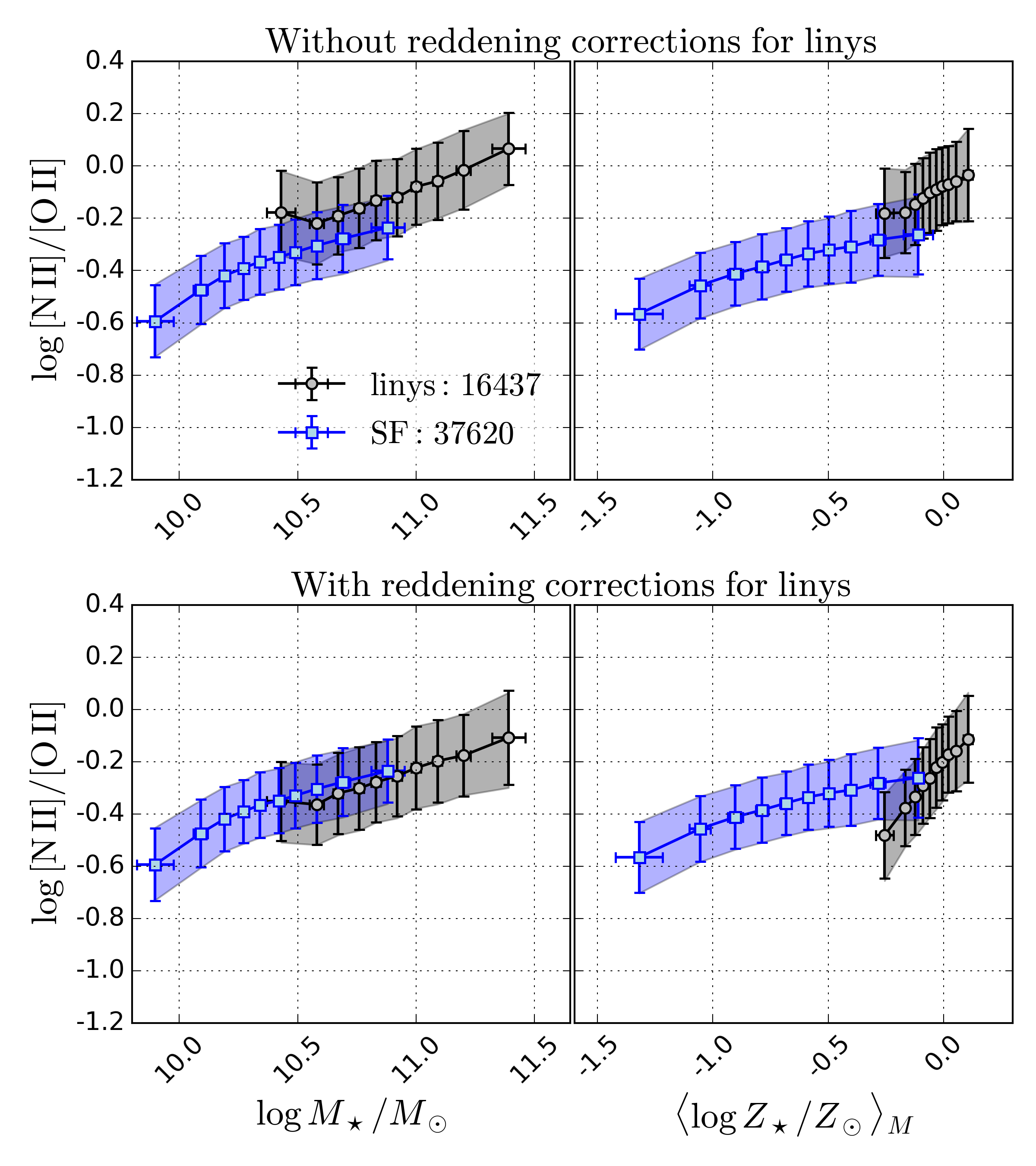}
\caption{
The ratio \nii/\oii as a function of the stellar mass (left panels) and of the metallicity (right panels) for star forming galaxies (blue) and liny RGs (black). The markers represent the median for bins containing the same number of objects. The shadowed regions are the quartiles of both distributions and the bars represent the median absolute deviation. The emission lines of liny RG are displayed without reddening correction (upper panels) and with correction using $5\times A_V$ (lower panels).
}
\label{fig:n2o2}
\end{figure*}

The chemical composition of the emitting gas can give a hint as of to its origin. The metallicity cannot be measured with confidence by any `strong-line method' because the physical conditions are certainly different from those for which the available methods were designed. One can instead use \nii/\oii as an indicator of the N/O ratio\footnote{The \nii/\oii ratio depends on N/O, on the metallicity and on the ionizing source.}. Since intermediate mass stars are the main source of nitrogen enrichment in massive galaxies \citep{Molla2015MNRAS451.3693M}, the gas from their stellar winds should be, on average, extremely nitrogen-rich. There is no accurate way to measure whether the emitting gas in liny RGs is nitrogen-rich or not, because the N/O ratio in the interstellar medium of galaxies increases with their metallicity. What we can do is to compare the \nii/\oii ratios of liny RGs with those of star-forming galaxies of same mass or of same stellar metallicity (as derived from \starlight)\footnote{The fact that the ionizing radiation field from HOLMES is harder than that of SF regions does have an impact on the \nii/\oii ratio, since the excitation potential of the \oii line is higher than that of the \nii line. But tests with photoionization models show that this effect is mild.}. This is done in Fig.~\ref{fig:n2o2}, where liny RGs are represented in black and SF galaxies\footnote{The SF galaxies entering this figure are chosen from a volume limited sample of the DR7-MGS with $0.04 < z < 0.095$ (similar as the volume limited sample in Section \ref{sec:datasamples}). Additional constraints are $SN > 3$ for \oii, \nii and \Ha. The SF galaxies are selected after \citetalias{CidFernandes2011}, with $\wha > 3$\,\AA\ and $\log\ \nii/\Ha < -0.4$.} in blue. The \nii/\oii ratios of SF galaxies are corrected for reddening using the observed \Ha/\Hb ratio. For liny RGs, the \Hb line is seldom available\footnote{In Fig.~\ref{fig:n2o2}, we considered only objects for which the signal-to-noise ratio of the relevant line intensities is larger than three.}. In the top panels, the \nii/\oii values are not reddening corrected, considering that the extinction of the stellar light as derived from \starlight is small (see Fig.~\ref{fig:AV}). It can be seen that the \nii/\oii ratios of liny RGs are comparable with those of SF galaxies of same masses or metallicities, being on average larger by about 0.1 dex.  If we consider that the nebular extinction in RGs is about five times larger than the stellar extinction, as obtained by \citet{Johansson2016MNRAS461.4505J} using stacked spectra, the intrinsic \nii/\oii ratios will be lower than they appear. In the bottom panels, we have corrected the observed values of \nii/\oii in RGs by multiplying the value of $A_V$ derived from \starlight by a factor of five. This  lowers the \nii/\oii ratios of liny RGs by about 0.2 dex. If the gas emitting the \nii and \oii lines originated from mass-loss of intermediate-mass stars, one would expect a difference in \nii/\oii similar to the one between planetary nebulae and \hii regions (i.e. about 0.7 dex from the data collected by e.g. \citealt{Bresolin2010MNRAS.404.1679B} or \citealt{Stasinska2013AA.552A.12S}). Thus we can definitely exclude that the emitting gas in liny RGs originates from stellar mass loss. 

This gas cannot come from mergers with low-metallicity galaxies either. Indeed, in Fig.~\ref{fig:n2o2},  liny RGs are superimposed on the SF galaxies in the zone of common stellar masses, and extend the pattern observed for the SF galaxies to higher stellar masses with a similar slope. Therefore, the ratio \nii/\oii in liny RGs is similar to that of rather massive SF galaxies and at least 0.5 dex larger than  in SF galaxies with masses smaller than $10^{9}$\msun\ as can be read from Fig.~\ref{fig:n2o2}. This indicates that the emitting gas in liny RGs does not come from the merger with a metal-poor galaxy.

\section{Discussion}\label{sec:discussion}

\begin{figure}
\centering
\includegraphics[width=.48 \textwidth]{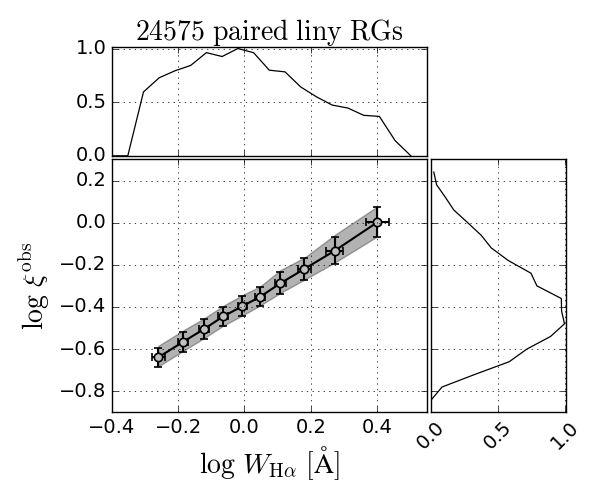}
\caption{
The equivalent width of \Ha versus $\xi$ for liny RGs from our paired
sample. The open circles are the median value of bins containing the same number of objects. The shadow region represent the quartiles of the distribution.
}
\label{fig:xiliny}
\end{figure}

The broad conclusion from the previous sections is that, after a careful matching in stellar mass, redshift and fibre covering fraction, we find that the HOLMES populations in liny RGs produce exactly the same number of ionizing photons as in the lineless ones (as shown by Fig.~ \ref{fig:qholmes}). We also note that the \Ha luminosity from liny RGs is compatible with the photons produced by HOLMES populations. Following \citetalias{CidFernandes2011}, we define $\xi = L_{\Ha}^\mathrm{obs}/L_{\Ha}^\mathrm{exp} (t > 10^8\,\mathrm{yr})$, where $L_{\Ha}^\mathrm{exp} (t > 10^8\,\mathrm{yr})$ is the \Ha luminosity expected from HOLMES. This parameter $\xi$ should be $\leqslant 1$ for galaxies whose ionizing spectra is powered by HOLMES (stellar populations with $t > 10^8$\,yr), while any other source of the ionization, such as star formation or an active nucleus should produce a value $\xi \gg 1$. Fig.~\ref{fig:xiliny} shows that liny RGs do not need any source of ionization other than HOLMES. Note that, as a matter of fact, the distribution of the values of $\xi$ peaks at log $\xi = -0.4$, implying that a significant fraction of the ionizing photons produced by the HOLMES actually escapes even from liny RGs \citep{Papaderos2013AA.555L.1P,Gomes2015arXiv151100744G}.

We have shown, however, that liny RGs differ from the lineless ones by having slightly higher values of the stellar extinction $A_V$. This is not surprising since lineless RGs having no warm gas are also expected to be devoid of dust. Liny RGs have higher luminosities in the WISE $W3$ band and higher luminosities in the GALEX $NUV$ band. Taken together, both facts point to a relatively recent period of star-formation in liny RGs, which does not occur in lineless RGs. This is also detectable in the diagram showing $D_n4000$ (Fig.~\ref{fig:d4000}) as well as in the one showing the mean luminosity-weighted age derived from \starlight (Fig.~\ref{fig:tstar}).

The gas out of which the stars formed and whose remnant is detectable through emission lines cannot come from stellar mass-loss in the old stellar populations, because, as seen in the previous section, it is not nitrogen-enriched. So it must have mainly an external origin. An additional argument for its origin could come from a comparison of the kinematics of the emission-line gas and that of the stars \citep[e.g.][]{Sarzi2006MNRAS.366.1151S,Davis2016MNRAS.457.272D}. The necessary information is not available in our SDSS data,  but in a sample of about 50 LIER galaxies with extended emission from the MaNGA survey, \citep{Belfiore2017MNRAS.466.2570B} found that the distribution of the star-gas misalignment indicates an external origin of the emitting gas (although they argue that internal processes may have a secondary role).

There are a number of studies that argue that a minor merger could be the cause of infrared and ultraviolet emission in many ETGs \citep[e.g.][]{Salim2010ApJ.714L.290S,Kaviraj2011MNRAS411,Sheen2016ApJ.827.32S}. However, as seen in the previous section, the \nii/\oii ratio of liny RGs indicates that the impacting galaxies cannot be low-metallicity SF galaxies.

The only remaining possibilities that we can think of is that the emitting gas  comes i) from accretion from the haloes of the galaxies, ii) from the intergalactic medium or iii) from residual streams of metal-rich gas coming from a merger in the recent past. 

In the case of accretion from the haloes, one would a priori expect the emitting gas to be enriched in products from stellar mass-loss and supernova explosions (the so-called wind-recycling studied by \citealt{vandeVoort2016MNRAS.462.778V}). However, the analysis of the chemical composition of galactic haloes using X-rays or far ultra-violet absorption spectra brought surprises and is not yet fully understood \citep{Pipino2011AA.530A.98P,Su2013ApJ.766.61S,Prochaska2017ApJ.837.169P}. Besides there is presently no measurement of the N/O ratio in halo gas (and no theoretical estimate of it either). Anyway, from the point of view of the chemical composition, accretion  from galactic haloes cannot presently be discarded to explain the emission lines in liny RGs. In the second scenario the gas in liny RGs would come from streams of cold gas found in the intergalactic medium, usually associated with galaxy groups or filaments \citep[e.g.][]{Sancisi2008AARv.15.189S,vandeVoort2012MNRAS.423.2991V}. This looks less probable, since the intergalactic medium gas is expected to be very metal poor, $\sim 0.1 Z_\odot$ \citet{Danforth2008ApJ.679.194D,Oppenheimer2012MNRAS.420.829O,vandeVoort2012MNRAS.423.2991V}, which is far below what our most metal-poor liny RGs suggest (see Fig.~\ref{fig:n2o2}). Finally, in the third scenario, cold gas comes from a merger episode with a metal-rich galaxy and now, a few billion years after the merger, it is slowly falling back to the galaxy. The first and third scenarios are the most probable for our liny RGs given the observed \nii/\oii ratios.

\begin{figure}
\centering
\includegraphics[width=.45\textwidth]{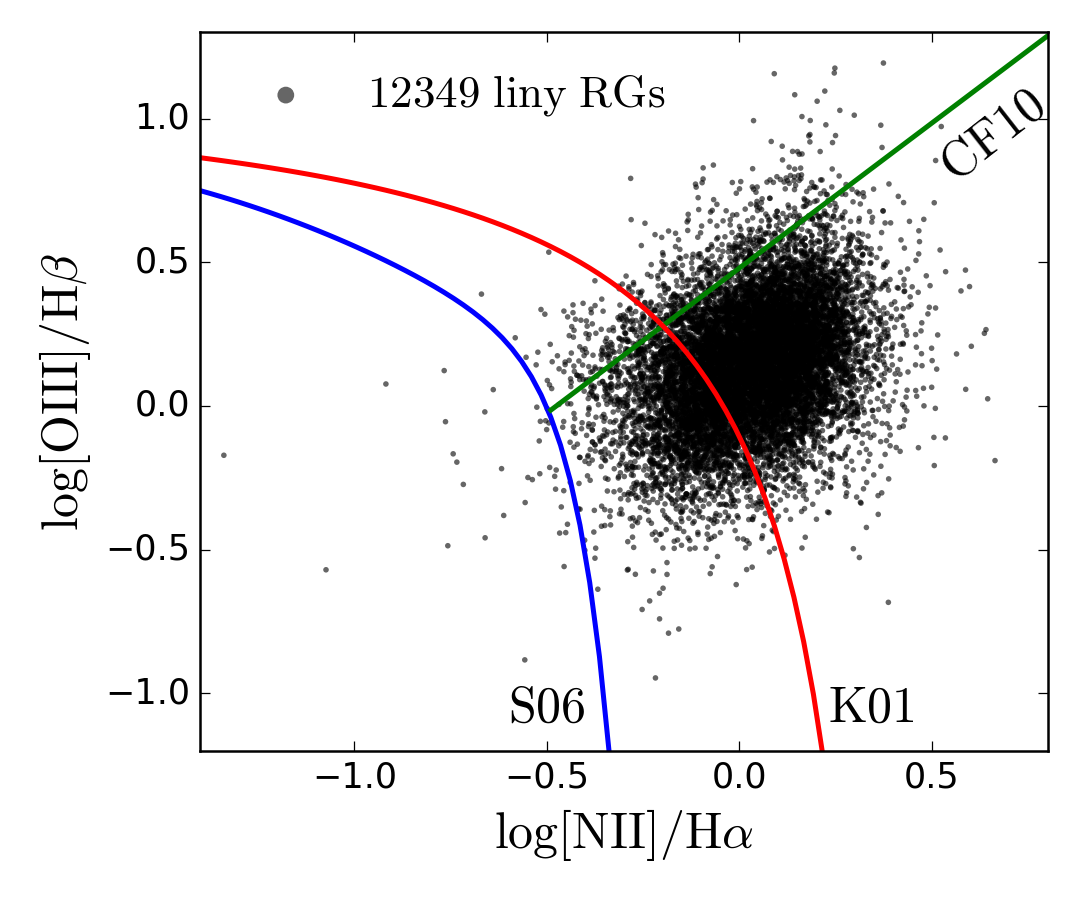}
\caption{
BPT diagram for the 12\,349 liny RGs
that survive  the four-line $SN$ cut. The blue line (S06) represents
the pure-SF separation from \citet{Stasinska2006}, the red one (K01)
the so called `upper starburst limit' from \citet{Kewley2001} and the green (CF10) is
the  Seyfert--LINER separation from \citet{CidFernandes2010}.
}
\label{fig:bpt}
\end{figure}

Our study indicates that liny RGs suffered an intermediate-age episode of star formation which is likely connected with the gas that is now being ionized by the HOLMES from the old stellar populations. Liny RGs \textit{are not} ionized by young stellar populations since in the BPT diagram  almost all are located well above the pure SF limit drawn by \citet{Stasinska2006}, as seen in Fig.~\ref{fig:bpt} which shows the values of \oiii/\Hb vs \nii/\Ha \citep{Baldwin1981} for our liny RGs. The figure plots only objects with $SN > 3$ for all four BPT emission lines, which drastically reduce the sample by a factor of 5. Only 0.7 percent o the RGs plotted in the figure are found inside the pure star-forming region. (Note that almost all the liny RGs live in the region previously attributed to LINER galaxies by \citealt{Kauffmann2003a}, \citealt{Kewley2006}, etc.). It must be noted, however, that in some early-type galaxies, massive ionizing stars may still be present, as indicated for example by 2D studies of early-type galaxies using CALIFA, which show zones where the \Ha equivalent width is of the order of 6-8 \AA, therefore not attributable to HOLMES \citep{Gomes2015arXiv151100744G}.

Lineless RGs, on the other hand, did not suffer such recent episodes of star formation, and are thus devoid of cold gas. In a companion paper (Mateus et al., in prep.), we will look for environmental clues for their lack of gas.

A major point of concern is what happens to the gas ejected by stellar winds, which should be extensively observed in all retired galaxies. The mass ejected back to the ISM is usually of the order of the galaxy stellar mass ($\sim 4\times10^{9} - 3\times10^{12}\,\mathrm{M}_\odot$ for our sample). However, calculations show that gas from stellar mass-loss \citep{Parriott2008ApJ681.1215P} and planetary nebulae \citep{Bregman2009ApJ699.923B} in ETGs is quickly heated to very high temperatures and it is not clear whether the amount of remaining warm gas is sufficient to explain the observed \Ha luminosities. 

One can then ask why the same does not occur with the infalling gas. The dynamics of infalling gas is very different. Accretion time-scales of infalling cold gas have been discussed by e.g. \citet{Davis2016MNRAS.457.272D} and argued to be long. The simulation of a test case of a massive ETG experiencing a major merger shows that a disk of gas misaligned with the stellar component is produced and persists for about 2 Gyr \citep{vandeVoort2015MNRAS.451.3269V}. Such conditions can allow the gas to live long enough for the emission lines be noticed.

\section{Summary}\label{sec:conclusions}

The main concern of this study was to find clues about why some RGs present emission lines and others do not. In this paper we compared the physical properties of  liny and lineless RG classes. 
In a volume-limited SDSS spectroscopic sample of ETGs, we pair-matched liny and lineless RGs in stellar mass, redshift and $R_{50}$ in order to avoid biases during the comparison procedure. We performed the comparison as a function of stellar mass and found the following.

\begin{enumerate}

\item There is no difference in the ionizing photon  budget of their old stellar populations ($Q_{\rm HOLMES}$). This means that both liny and lineless RGs have HOLMES capable of producing the same amount of \Ha emission, so the difference between those two classes must lie in their warm gas content.

\item There is a systematic difference in $D_n4000$ and luminosity weighted mean stellar ages, indicating that liny RGs are younger on average, or have a younger component. Those differences are very small to be reliable on their own. Their significance is however reinforced by the excess $NUV$ in liny RGs. This means that the small mean stellar age difference is not due to very young stars, but rather to intermediate-age stellar populations (0.1--5 Gyr).

\item Liny RGs have higher $A_V$ and $W3$, which indicates either a difference in dust content or an enhanced population of HOLMES heating their expelled dust shells.

\item The amount of gas responsible for the \Ha emission in liny RGs is smaller by orders of magnitude than the amount of matter ejected by winds from intermediate-mass stars. In other words, mass-loss from intermediate-mass stars would provide enough material to explain the \Ha emission. 

\item The \nii/\oii emission-line ratio in liny RGs rules out the hypothesis that the line-emitting gas in RGs comes from stellar winds. It also rules out the hypothesis of metal-poor minor mergers being at the origin of this gas.

\end{enumerate}

The overall conclusion of our work is that liny RGs must have experienced in the recent past an episode of star formation, and that the origin of the emission-line gas is either gas coming from the haloes of RGs (such as gas ejected in strong winds by massive OB stars or an AGN), or the left-over gas around the galaxy after the merger episode. The ionization of that gas would be due to HOLMES. On the other hand, lineless RGs would be RGs that did not experience such a recent merger or inflow of chemically enriched material.
 
We however note a problem: the hypothesis of a wet merger with a high-metallicity galaxy seems at odds with the current idea that ETGs in the local universe tend to grow via minor and gas-poor mergers \citep{Naab2009ApJ.699L.178N,Conselice2014ARAA.52.291C,Kaviraj2014MNRAS.437L.41K}. It is not clear whether this excludes the possibility of one merger with a metal-rich star-forming galaxy, such as modelled by \citet{diMatteo2007AA.468.61D}. The hypothesis of enriched gas accreting from the haloes of the galaxies seems a priori less problematic. Simulations by \citet{Oppenheimer2010MNRAS.406.2325O} show that late-time gas accretion and the resulting star formation is due to material previously expelled from the galaxy. However, as pointed out by those authors themselves and also by \citet{vandeVoort2016MNRAS.462.778V}, such results depend on many issues not yet fully treated by the simulations.

A remaining question is what happens to the gas returned to the ISM by old stellar populations. In a typical massive early-type galaxy, the amount of ejected gas is about the same as its present-day stellar mass. Where does this gas go? There are a number of mechanisms which can explain the absence of warm gas in the ISM, usually involving a short time scale ($t_d \sim 10^5$\,yr) for gas destruction or heating. Kinematically-disconnected gas, such as externally accreted gas, can survive longer and thus be ionized by HOLMES, emitting emission lines which embellish a liny RG's spectrum.

In a companion paper (Mateus et al. 2018, in prep.) we will present environmental factors linked to the presence or absence of emission lines in RGs, and discuss some possible scenarios that can explain the observations.

\section*{Acknowledgements}
We thank the anonymous referee for the numerous suggestions that helped improve this work. F. R. Herpich thanks FAPESC and CAPES for the financial support during this work.  The authors acknowledge the support from the CAPES CSF--PVE project 88881.068116/2014-01. NVA acknowledges support of the Royal Society and the Newton Fund via the award of a Royal Society--Newton Advanced Fellowship (grant NAF\textbackslash{}R1\textbackslash{}180403), and of FAPESC and CNPq.
The Sloan Digital Sky Survey is a joint project of The University of Chicago, Fermilab, the Institute for Advanced Study, the Japan Participation Group, the Johns Hopkins University, the Los Alamos National Laboratory, the Max-Planck-Institute for Astronomy, the Max-Planck-Institute for Astrophysics, New Mexico State University, Princeton University, the United States Naval Observatory, and the University of Washington.  Funding for the project has been provided by the Alfred P. Sloan Foundation, the Participating Institutions, the National Aeronautics and Space Administration, the National Science Foundation, the U.S. Department of Energy, the Japanese Monbukagakusho, and the Max Planck Society.

\bibliographystyle{mnras}
\bibliography{library}

\bsp
\label{lastpage}

\appendix
\section{The pairing lineless $vs$ liny}\label{app:pairing}

To guarantee a reliable description of the physical properties involved in our analysis, we draw a liny RG sample pair-matched to our lineness RG sample.  To do this, we match the samples in three physical properties: the galaxy mass from \starlight ($M_\star$), the redshift ($z$), and the Petrosian radius that contains half of the light of the galaxy in the $r$ band ($R_{50}$). We use these parameters to find the most structurally similar pairs in our liny and lineless RG samples. Each one of these three properties is a parameter $P$ that will be used to calculate the similarity between a lineless and a liny galaxy.

The methodology of the pair-matching procedure is quite simple: given a lineless RG galaxy, we compare each parameter $P$ with the same parameter of those objects in the liny RG group. At the end, the liny galaxy with the smallest cumulative difference is chosen. Note that each lineless galaxy will be unique but the same is not necessarily true for liny RGs, as a single liny RG can be an ideal match to more than one lineless galaxy. Quantitatively, we compute the absolute value of the difference between the parameters $P$ for a pair of galaxies, and normalize it by the standard deviation of $P$ ($\sigma_P$) for all RGs in our sample. Finally, we sum the differences for all parameters $P$ to obtain the similarity parameter $SP$: \begin{equation}\label{eq:similarity}
     SP_j = \sum_P \frac{|P^{\mathrm{lineless}} - P^{\mathrm{liny}}_{j}|}{\sigma_P},
\end{equation}
where $P = \{\log M_{\star}, z, R_{50}\}$, $j = 1, 2, 3, \dots, N$, and $N$ is the total number of liny RGs.

We apply this procedure to our sample of 96\,844 lineless and 59\,662 liny RGs, described in Section~\ref{sec:sample}. After this, each lineless at the list will have one ideal liny partner that minimizes $SP$. We then exclude outliers from the sample, choosing only pairs in which the value $|P^{\mathrm{lineless}} - P^{\mathrm{liny}}|$ is smaller than $3 \sigma_P$, where $\sigma_{\log M_\star} = 0.357$\,dex, $\sigma_z = 0.054$ and $\sigma_{R_{50}} = 2.493$\,kpc, which excludes 8 pairs. Our paired sample is comprised of 96\,836 parent lineless RGs that are paired with 41\,286 unique liny RGs. From the pair-matched liny RGs, 17\,617 are matched to only one lineless RG, 10\,853 are matched to 2 lineless RGs, 5948 to 3, 2947 to 4, 1532 to 5, 2021 to 6--10, and 368 to more than 10 lineless RGs. The maximum number of lineless parents for a single liny RG is 103.

From those pairs, we select our volume-limited samples. The optical VLS sample has 24\,575 lineless--liny pairs containing 13\,463 unique liny RGs; the WVLS has 6944 pairs containing 4838 unique liny RGs; and the GVLS sample has 8121 lineless--liny pairs containing 5644 unique liny RGs.

\end{document}